\documentclass[12pt, onecolumn, draftclsnofoot]{IEEEtran}
\usepackage{times}
\usepackage{amsmath,epsfig}
\usepackage{amssymb}
\usepackage{amsfonts}
\usepackage{array,dcolumn}
\usepackage[ruled,vlined,linesnumbered]{algorithm2e}
\usepackage{subcaption}
\usepackage{psfrag}
\usepackage{amsmath}
\usepackage{amsfonts}
\usepackage{amssymb}
\usepackage{balance}
\usepackage{accents}
\usepackage{booktabs, adjustbox}
\usepackage{cite}
\usepackage{ragged2e}
\usepackage[all]{xy}
\usepackage{dsfont}
\usepackage{hyperref}
\usepackage{url}
\usepackage{xcolor}
\usepackage[nolist]{acronym}
\usepackage{graphicx} 
\usepackage{float}
\usepackage{threeparttable}
\usepackage{cancel}
\usepackage[official]{eurosym}
\usepackage{subcaption}
\usepackage{fancyhdr}
\usepackage{comment}
\usepackage{caption}
\usepackage{subcaption}
\usepackage{chemformula}
\hypersetup{
     colorlinks = true,
     linkcolor = blue,
     anchorcolor = black,
     citecolor = red,
     filecolor = blue,
     urlcolor = blue
     }

\newcommand{\PP}[1]{\textcolor{black}{#1}}

\newcommand{\AL}[1]{\textcolor{black}{#1}}

\newcommand{\ilm}[1]{\textcolor{black}{#1}}

\newcommand{\dd}{\mathop{}\!\mathrm{d}}

\title{B-ETS: A Trusted Blockchain-based Emissions Trading System for Vehicle-to-Vehicle Networks}

\author{\IEEEauthorblockN{Lam Duc Nguyen$^1$, Amari N. Lewis$^2$, Israel Leyva-Mayorga$^1$, \\ Amelia Regan$^2$, and Petar Popovski$^1$}\\
\IEEEauthorblockA{$^1$Department of Electronic Systems, Aalborg University, Denmark \\ 
$^2$Department of Computer Science, University of California, Irvine, United States}\\
\IEEEauthorblockA{E-mail: \{ndl, ilm, petarp\}@es.aau.dk, \{amaril, aregan \}@uci.edu }}

\begin{document}
\maketitle

\begin{abstract}
Urban areas are negatively impacted by Carbon Dioxide (\ch{CO_2}) and Nitrogen Oxide (NO\textsubscript{x}) emissions. In order to achieve a cost-effective reduction of greenhouse gas emissions and to combat climate change, the European Union (EU) introduced an Emissions Trading System (ETS) where organizations can buy or receive emission allowances as needed.
%
The current ETS is a centralized \PP{one, consisting of a set of  complex rules. It is currently administered at the organizational level and is used for fixed-point sources of pollution such as factories, power plants, and refineries. However, the current ETS cannot efficiently cope with vehicle mobility, even though vehicles are one of the primary sources of \ch{CO_2} and NO\textsubscript{x} emissions.}
In this study, we propose a new distributed Blockchain-based emissions allowance trading system called B-ETS. \PP{This system enables transparent and trustworthy data exchange as well as trading of allowances among vehicles, relying on vehicle-to-vehicle communication.} 
In addition, we introduce an economic incentive-based mechanism that \PP{appeals to} individual drivers and leads them to modify their driving behavior \PP{in order} to reduce emissions. 
\PP{The efficiency of the proposed system is studied through extensive simulations, showing how increased vehicle connectivity can lead to reduction of the emissions generated from those vehicles.} We demonstrate that our method can be used for full life-cycle monitoring and fuel economy reporting. \PP{This leads us to conjecture that the proposed system  could lead to important behavioural changes among the drivers.}
\end{abstract}



\section{ \uppercase{Introduction}}
\begin{figure}
    \centering
    \includegraphics[width=0.7\linewidth]{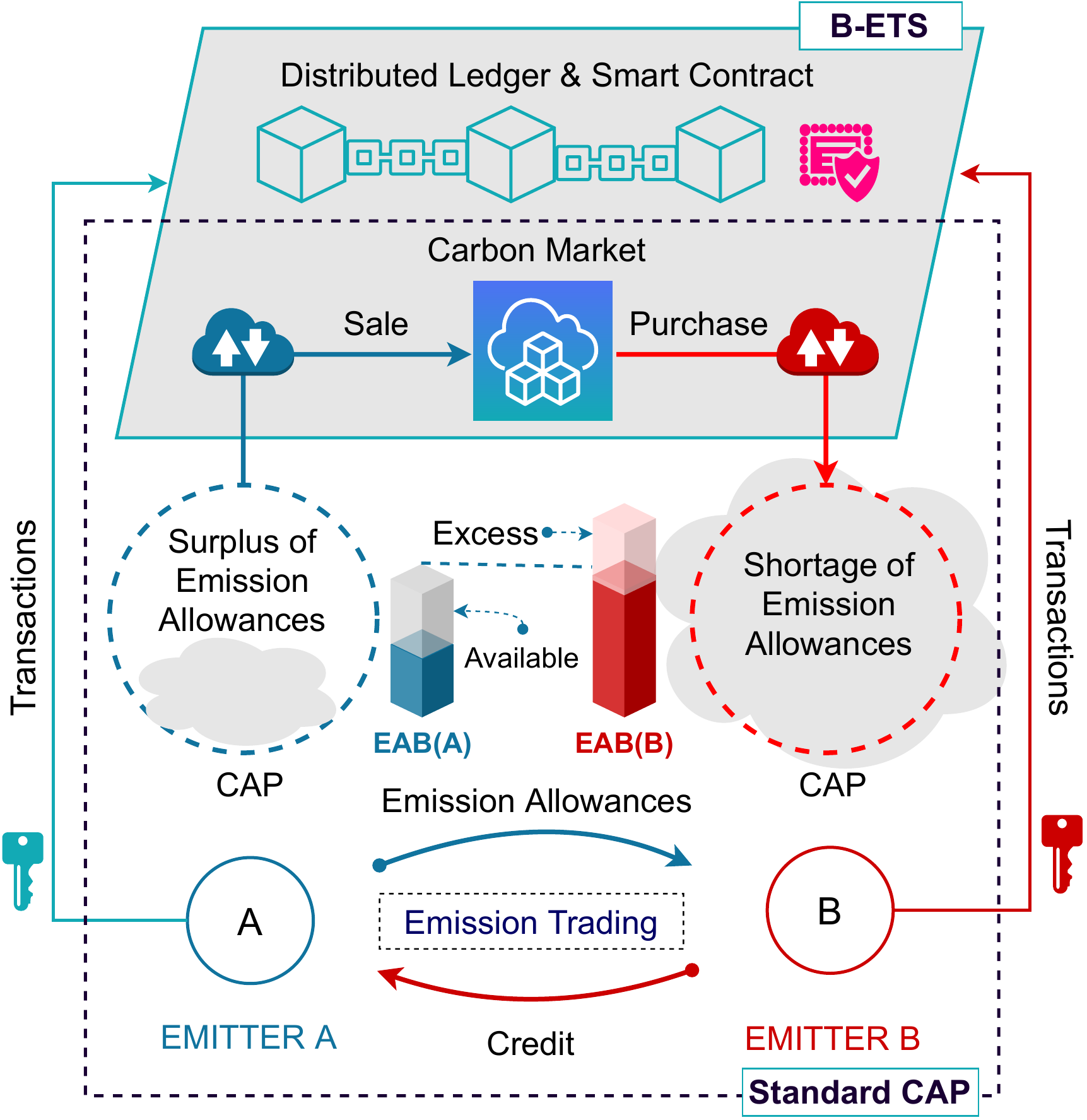}
    \caption{B-ETS general architecture.}
    \label{fig:fig1overview}
\end{figure}

\noindent Typical passenger vehicles emit about 4.6 metric tons of carbon dioxide \ch{CO_2} per year. 
The European Union's Emission Trading System (EU-ETS) is the world's first major carbon trading market with the main goal to combat climate change and reduce Greenhouse Gas (GHG) emissions in a cost effective way.
The EU-ETS works on a Cap-and-Trade (CAP) principle which allows companies that generate point source emissions to receive or buy emission allowances, which can be traded as needed ~\cite{commision2015road}. 
\PP{The process of \AL{our B-ETS} CAP program is described in \AL{Figure~\ref{fig:fig1overview}}, where it is seen that it is based on a complex centralized method of trading among the organizations involved. The first step in CAP is 
to make a centralized decision} (by a regulatory agency or some other collective entity) on the aggregate quantity of emissions allowed. Allowances are then written in accordance with this quantity, \PP{after which they are distributed among the sources responsible for the emissions.} 

Since 2018, the EU-ETS began penalizing vehicle manufacturers for exceeding the targets for fleet-wide emissions for new vehicles sold in any given year. The manufacturers are required to pay an excess emissions premium for each newly registered car. A penalty of \euro{95} must be paid for each gram per km above the target \cite{commision2015road} and the target of \ch{CO_2} for the 2020-2021 period is set to 95 grams per km. 
In this work, we address the need for a new trusted and distributed system which can audit emissions at the vehicle-level. 

The emerging Distributed Ledger Technologies (DLTs) brought a new era of distributed peer-to-peer applications and guarantees trust among involved parties. The terms DLT and Blockchain will be used interchangeably throughout this paper, Blockchains are a type of DLTs, where chains of blocks are made up of digital pieces of information called transactions and every node maintains a copy of the ledger. In DLTs, the authentication process relies on consensus among multiple nodes in the network \cite{LamNames}. Each record has a timestamp and cryptographic signature; the system is secure and maintains a transaction ledger that is immutable and traceable. Ultimately, the goal of applying Blockchain technology to the transportation industry is to provide a fully distributed ETS system that can encourage direct communication between producers and consumers. A primary reason to embrace new DLTs is to bypass the administrative pitfalls that have plagued current emissions monitoring systems. Security is another aspect that motivates this approach. For instance, data pollution attacks are incredibly dangerous, these attacks typically occur in centralized systems and involve an adversary trying to modify the content of the packets and then forward the corrupted messages to neighboring nodes. The integration of Blockchain in individual carbon trading will accelerate the involvement of the public in carbon trading and sensitize society to individual level carbon footprints.

Current V2V approaches have limitations such as: the need for trusted third-party entities, security hardware, higher communication and storage overhead, high implementation costs, and issues related to the confidentiality of data. Studies \cite{zheng2017influence}, \cite{alsabaan2013optimization}, \cite{li2018creditcoin} have strictly considered Vehicle-to-Infrastructure (V2I) approaches incorporating additional resources such as On-board Units (OBUs) and Roadside Units (RSUs). 
Eckert et al. develop a carbon Blockchain framework for Smart Mobility Data-Market as a trading system for \ch{CO_2} in  the form of carbon tokens in \cite{Eckert}. The evaluation is done on the user and vehicular levels. 
Pan et al. outlined some advantages of the use of Blockchain in ETS namely safety and reliability, efficiency, convenience, openness and inclusiveness \cite{Pan}. That work was not concerned with V2V networks or mobile carbon emissions trading, but it did introduce the concept of personal carbon emissions trading which could be applied in vehicular networks. 

  In this study, we first tackle the challenges of the current EU-ETS system by proposing a distributed emissions allowance trading system called B-ETS. The system creates an account for the emissions generated from each vehicle and allows exchanges among vehicles in a trusted manner based on Blockchain and Smart Contracts.
In B-ETS, each vehicle acts as a light client in the global Blockchain network and manages its own Emission Allowance Balance (EAB) which is reset at the beginning of each day. The EAB data is recorded transparently and immutably in the distributed ledger. \PP{It should be noted that} we use one day as our unit of time without loss of generality. Any other unit (a week, a month) could be used if that seemed more suitable.

  Then, we introduce an economic incentive-based mechanism which attracts drivers to change their driving behavior in order to reduce emissions. Each vehicle's generated emissions are calculated and the data are recorded immutably in the distributed ledger. If the emission level is higher than the defined threshold, the EAB will be reduced. If the EAB goes to zero, the driver needs to buy credits in the form of EAB from others. 

  The proposed V2V-based allowance trading system would not replace the in-service fleet-wide monitoring required by the EU-ETS plan. Rather, it would complement that plan by making it the responsibility of drivers to meet personal emissions targets. \ilm{That is, without individualized feedback, drivers cannot measure the environmental impacts of their actions. Furthermore, without incentives, they might not be willing to contribute to environmental sustainability.} 
 
  \AL{Given the proposed B-ETS system, vehicles participating in the program will be influenced by the economic incentive. Drivers are more prone to behave better when their EAB and driving privileges are at stake. If drivers contribute to lower emissions (i.e., demonstrate healthy driving habits), their EAB will increase or remain positive. Essentially, drivers want to avoid having to purchase credits from others or having a negative EAB balance as this could lead to driving restrictions.}
  
 \AL{ Our mechanism can be compared to the traffic point penalty system in the U.S., Canada and other countries. As punishment for committing traffic violations, the drivers risk the suspension  or revocation of their license based on a point-record mechanism in place. As a result, the Department of Motor Vehicles (DMV) can revoke the driver's license of that person and they are not allowed to drive any motor vehicle. In order to mitigate the social cost of license suspensions, point-removal systems exist for most point-record drivers licenses \cite{dionne2011incentive}. In contrast, our system proposes a daily (or weekly or other period as appropriate) record of associated driving behaviors with vehicle emissions data and individual accounts.}
  
The execution of the smart contract guarantees trust among vehicles and driving habits, (e.g, avoid idling, speeding, etc) and \ch{CO2} levels. Vehicles in the system are alerted via rules defined in the smart contract to reduce emissions \cite{Dolowers13:online} \cite{zheng2017influence}. 

  Our solution to reducing vehicle \ch{CO2} emissions involves the use of DLT-enabled emissions monitoring, which could be applicable to any market worldwide. In this work, we focus on the EU, but, our method can comply with regulations in China and could be implemented in the US to measure life-cycle Corporate Average Fuel Efficiency (CAFE) standards.

  The contributions of this study are described as follows:

\begin{itemize}
    \item First, we propose a distributed Blockchain-based emission trading system named B-ETS that will meet the requirements of the EU-ETS plan for reducing vehicular emissions. B-ETS overcomes the disadvantages of current centralized ETS systems and provides a trustworthy approach for exchanging data in vehicle-to-vehicle networks.
    \item Second, we introduce an economic incentive-based system which motivates drivers to reduce fuel consumption and pollution. Based on the autonomous execution of smart contracts, the incentive mechanism is guaranteed to work in a trusted and distributed manner. 
    \item Third, realizing the lack of communication and computation analysis in Blockchain-enabled vehicle networks, we present a theoretical model to derive the communication \PP{efficiency} of the proposed system B-ETS. 
\end{itemize}
The remainder of this paper is organized as follows. In the next section, we present the system model and analysis. In section III, the performance evaluation is outlined including our results. Finally, in section IV, we provide our conclusion and plan future work.

\section{ \uppercase{System Model and Analysis}}
\subsection{Blockchain as a Ledger for VANET}
The system operates within periods of duration $T$. In this section, we describe the two major system components: the vehicles and the distributed ledger, followed by the selected model for \ch{CO2} emissions. Table~\ref{tab:symnonym} presents the nomenclature used throughout the paper.

\begin{table}[t!]
\centering
\caption{\ilm{Nomenclature}}
 \begin{tabular}{@{}  p{1.2cm}  p{5.8cm} @{}}
 \toprule
 \textbf{Symbols} & \textbf{Descriptions} \\ [0.5ex] 
 \midrule
 $T$ & Considered system period [hours] \\
$\mathcal{V}$ & Set of vehicles \\
$i$ & Vehicle $i\in\mathcal{V}$\\
$T_s$ & \ch{CO2} sampling period\\
$\epsilon_i(t)$ & Average \ch{CO2} emissions per km for\\
& vehicle $i$ at time $t$\\
$B_i(t)$ & Emission allowance balance of \\
& vehicle $i$ at time $t\in\left[0,T\right)$\\
$p_i(t) $ & Penalty/tax for vehicle $i$ at time $t$  \\
$s_i(t) $ & Incentive (subsidy) for vehicle $i$ at time $t$ \\
$ L_{total} $ & Total allowed latency  \\
$ L_{trans} $ & Communication latency  \\
$ L_{comp} $ & Blockchain verification latency  \\
$R$ & Communication data rate [packets/s]  \\
$\mathbf{v}_i(t)$ & Speed of vehicle $i$ at time $t$ [km/h] \\
$ S_B $ & Blockchain block size in bits \\
$ \mathbf{v}_{ij}(t) $ & Relative speed between $i$ and $j$ at time $t$\\
$r_{ij}$ & Communication Range between $i$ and $j$\\
$e_{i,j}(t)$  & Allowances sold by $j$ to $i$ at time $t$ \\
$\mathcal{T}$ & Maximum allowed \ch{CO2} emissions generated by vehicles per km. \\
\bottomrule
\end{tabular}
\label{tab:symnonym}
\end{table}

\subsubsection{Vehicles} 
Let $\mathcal{V}$ be the set of vehicles in the system. An On-Board-Unit (OBU) is installed in each vehicle $i\in\mathcal{V}$ in the Blockchain-based VANET. The OBU performs light tasks, including collection and transmission data to other vehicles according to the IEEE 802.11p communication standard, and provides support to passengers and drivers. 

Within each system period of duration $T$, the \ch{CO2} emission monitoring system takes samples of the average \ch{CO2} emissions per km in each vehicle $i$ and updates the ledger. The \ch{CO2} is sampled at fixed intervals of duration $T_s<T$\,hours. The sample taken by vehicle $i$ at time $t\in\left\{0,T_s,2T_s, \dotsc,T\right\}$\,hours is denoted as $\epsilon_i(t)$ and consists of the taken measurement, the vehicle ID $i$, and a timestamp, generated as a function of $t$. 
The amount of \ch{CO2} generated at the vehicles is reset to zero at the beginning of each period of duration $T$, hence, $\epsilon_i(0)=0$. 

\begin{figure}[t]
    \centering
    \includegraphics[width=0.7\linewidth]{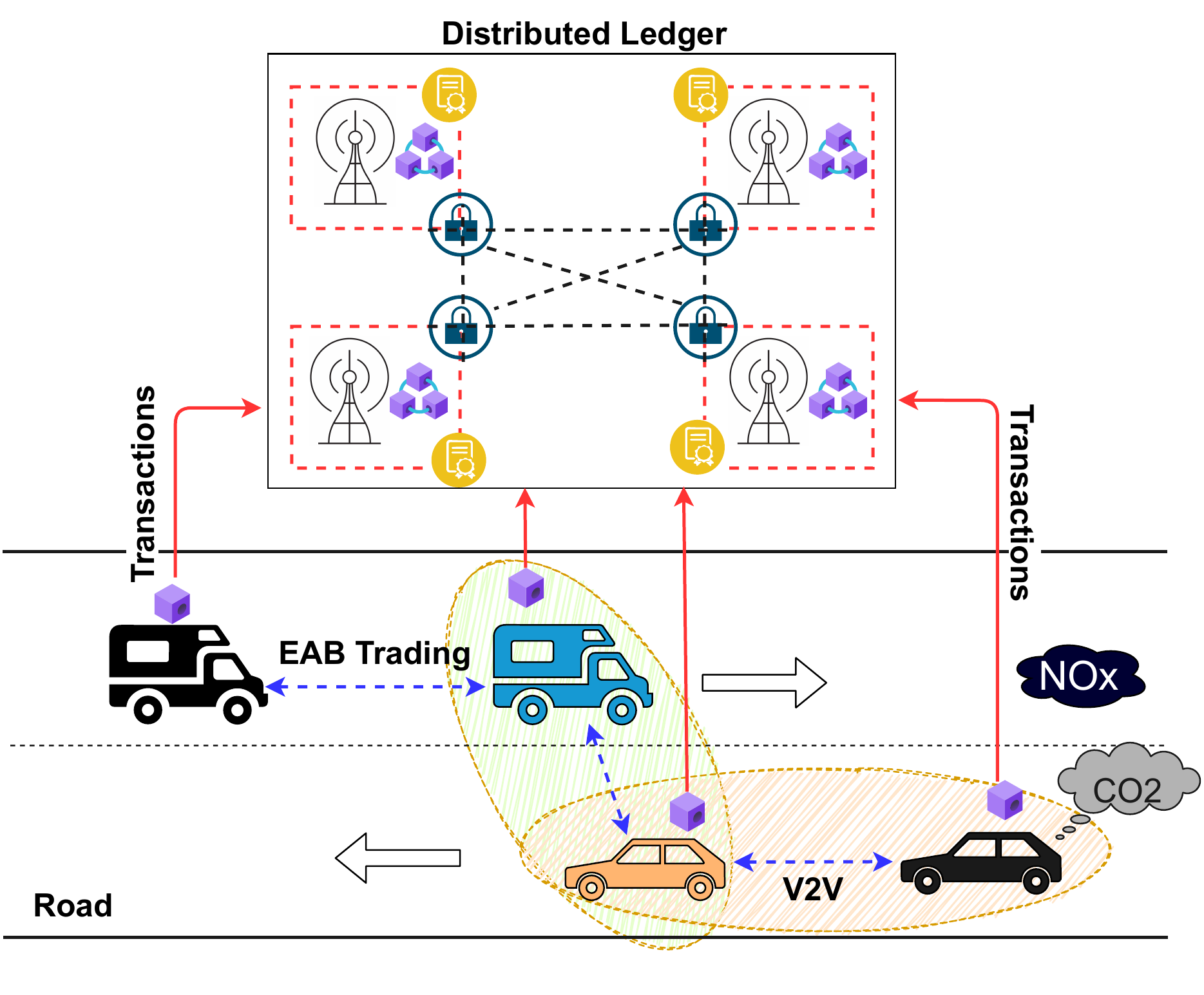}
    \caption{Blockchain-enabled vehicular emission trading system. }
    \label{fig2systemmodel}
\end{figure}

\subsubsection{Distributed Ledger} The distributed ledger records the data exchange history grouped into blocks and linked together chronologically. To minimize the cost of storage, the sensing data could be hashed and stored at more powerful nodes, and only the hash of data is recorded to the blockchain. Next, a confirmation message is sent back to confirm that the data has been added to the ledger \AL{as presented in Figure \ref{fig:workflow}}. We assume that the data services (e.g., data storage, trading and task dispatching) are implemented on top of a permissionless Blockchain \cite{9324804}.

\AL{In a permissionless blockchain, any peer can join and leave the network at any time as a reader or writer. Permissionless Blockchains are open and decentralized with no central authority. Bitcoin and Ethereum are instances of permissionless Blockchains. In contrast, in permissioned Blockchains a central authority decides and attributes the right to individual peers to participate in the write or read operations of the blockchain. Examples of these include Hyperledger Fabric and R3 Corda \cite{Crypto}.} 

The sensing data are formatted into transactions of fixed size. To enhance efficiency, only the digest of each transaction is stored on the chain, and the content of the transactions are stored by each consensus node off-chain or at the IPFS storage.

\subsubsection{Emissions Model} The amount of \ch{CO2} generated from vehicles depends on various factors such as: national average age distributions, vehicle activity speeds, operating modes, vehicle-miles traveled, starts and idling, temperatures, maintenance, anti-tampering programs, and average gasoline fuel properties in that calendar year \cite{Estimate62:online}. The calculation of emissions in our simulations are based on the Handbook Emission Factors for Road Transport V3.1 (HBEFA), the model was implemented by extracting the data from HBEFA and fitting them to a continuous function obtained by simplifying the function of the power the vehicle engine must produce to overcome the driving resistance force \cite{SUMO1}. %

\subsection{Emission Allowances Trading}%
\subsubsection{Traditional Cap-and-Trade}%
\AL{Traditionally, cap-and-trade commonly refers to governmental regulations and programs in place to limit the levels of \ch{CO2} emissions as a result of industry activity. As briefly mentioned, the EU-ETS works on a cap and trade principle, where the cap is a dynamic limitation, set on the total amount of GHG emitted by installations covered by the system. Within the system, companies receive or buy emission allowances which can be traded. Although, vehicular emissions were not initially considered, in 2006, researchers at MIT joint program on the science and policy of global change introduced the implementation of a cap-and-trade policy for vehicles. Their central conclusion indicated that there are important efficiency gains to be realized by including transport emissions under the CAP and by integrating pre-existing programs, such as CAFE, and cap-and-trade systems \cite{ellerman2006bringing}.}

\subsubsection{B-ETS Framework}
Our B-ETS framework considers an economy where vehicles produce goods over a system period $[0, T]$\,hours.
Therefore, each vehicle $i$ acts as a wallet in the Blockchain network and its EAB at time $t$ is denoted as $B_{i}(t)\in\mathbb{R}$. In the system, the updates to the EAB are triggered by the sampling of the \ch{CO2} emissions of the vehicles, hence, the system operates at specific times $t\in\{0,T_s,2T_s,\dotsc\}$.  
At the beginning of each period of duration $T$, the EAB of each vehicle $i$ is reset to a pre-defined value $B_{i}(0)$. So, the EAB cannot be accumulated between subsequent periods. However, if $i$ were to hold on to this initial allowance endowment until the end of the period, it would be able to offset the system's \textit{cap} by up to $B_{i}(0)$ units of emissions credits. This is the \textit{cap} aspect in our B-ETS scheme.

The EAB pertains to an individual account in which the allowances are used and exchanged amongst vehicles for environmental sustainability. In order to offset penalties, the vehicles with low balances may engage in buying allowances from vehicles that expect to meet demand with fewer emissions than their own cap. This is our \textit{trade} aspect of B-ETS framework. 

 \textbf{Remark 1}: \textit{A CAP program is only feasible in scenarios where the vehicles have a positive allowance balance at the beginning of the periods. Hence, the following inequality must hold:}
\begin{equation}
    B_i(0) >0\qquad \text{for all } i\in\mathcal{V}.
\end{equation}

\begin{figure}
    \centering
    \includegraphics[width=0.7\linewidth]{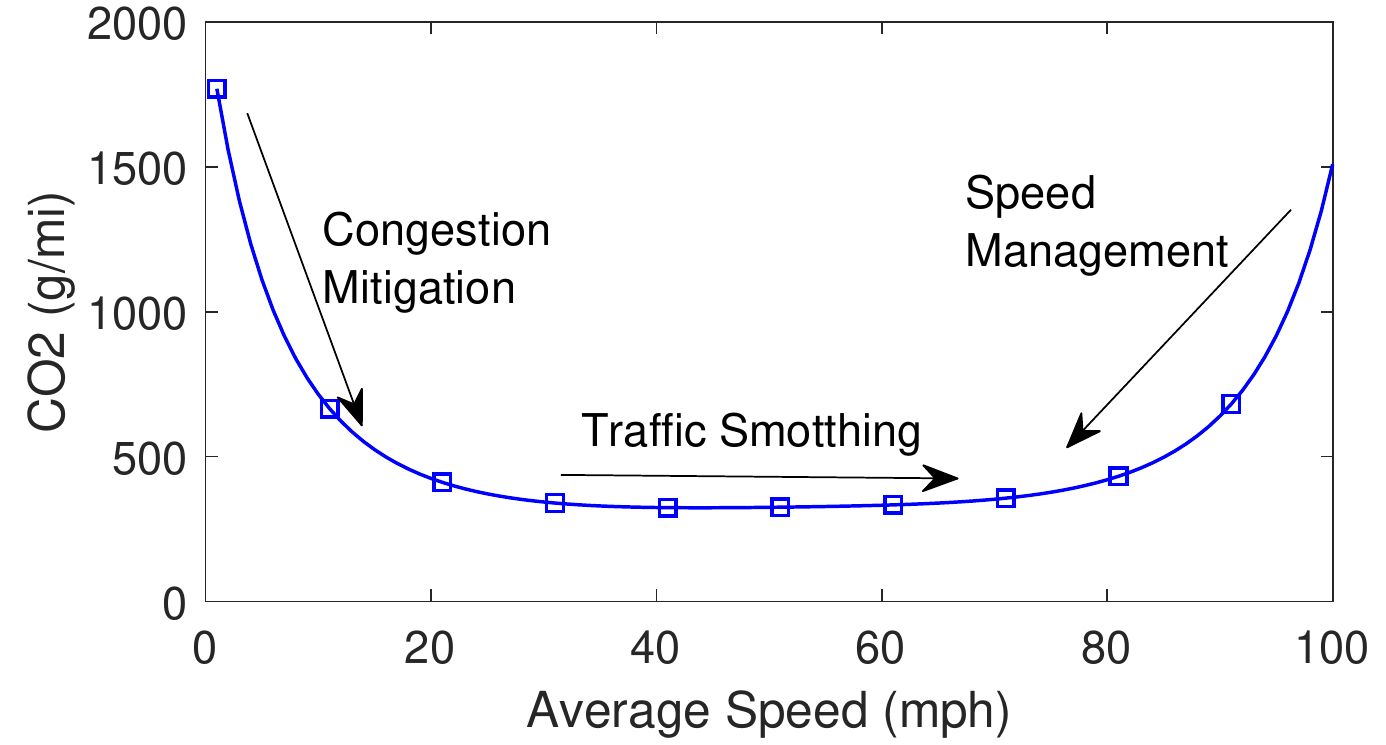}
    \caption{\ch{CO2} emissions (grams/mile) as a function of average speed (mph) \PP{\cite{cappiello2002statistical}}}
    \label{fig:co2speed}
\end{figure}

  The maximum allowed \ch{CO2} emissions generated by vehicles per km is denoted as $\mathcal{T}$ (which is defined as a rule in smart contract). If $\epsilon_i(t) > \mathcal{T}$, then our initial smart contract is executed to generate an alert to $i$ to reduce vehicle speed as a direct solution to reduce amount of generated \ch{CO2}, and the fine $p_{i}(t)$ will be deducted from its EAB. In contrast, the subsidy $s_i(t)$ will be endowed to $i$ for maintaining the \ch{CO2} emissions below $\mathcal{T}$. 

The values of $p_i(t)$ and $s_i(t)$ are considered as taxes and subsidies for vehicle $i$ that depend on their behavior. The incentive may help encourage the driver to control their driving behavior to avoid generating \ch{CO2} higher than the allowed standard. The driver needs to choose between receiving an incentive by reducing amount of emissions or being fined due to overloaded generated emissions. The penalties and subsidies are computed based on the theoretical model presented in \cite{Tax} which depends on various vehicular factors. 

In order to increase the subsidies and reduce the penalties, the drivers can follow strategies defined in smart contracts. For example, Figure \ref{fig:co2speed} shows that \ch{CO2} is  a function of average speed. First, we observe that very low average speeds generally represent stop and \AL{start} driving periods, and vehicles traveling in short distances, in these cases, the emission rates are quite high. In this period, the smart contract defines rules to increase traffic speeds and reduce congestion by, for instance avoiding high traffic roads to reduce emissions. Second, when the speed of the vehicle is too high, it demands high engine loads which require more fuel, leading to higher \ch{CO2} emission rates. The techniques to manage high speeds are implemented in the contracts which recommends the drivers to simply reduce their speeds. Consequently, moderate speeds of around 40 to 60 mph  are ideal speeds which reduce emissions and will give the drivers incentive to improve their balances.  

In addition, the EAB can be traded among vehicles based on predefined smart contracts. Whenever $B_i(t)<0$, there will be a red alert issued to $i$ for having a negative-balance. This alert is in the form of penalties, or restricted road access to zero-balance vehicles. In this cases, the vehicles can either wait until the next period for their EAB of to be reset or buy the EAB from other vehicles. We consider the case of vehicles exchanging EAB on-road via execution of smart contract and distributed ledger. For this, let $e_{i,j}(t)$ be the amount of allowances sold by vehicle $j$ from vehicle $i$ at time $t$. These operations are recorded in the distributed ledger.

\textbf{Remark 2}: \textit{The vehicle $j$ cannot sell more allowances $e_{ij}(t)$ than it actually owns. In other words, $i$ cannot buy more than is actually available. Hence,}
\begin{equation}
    e_{i,j}(t) \leq B_{j}(t), \text{for all } j\in\mathcal{V},\,t\in\left[0,T\right)
\end{equation}

\begin{figure}[t!]
    \centering
    \includegraphics[width=0.7\linewidth]{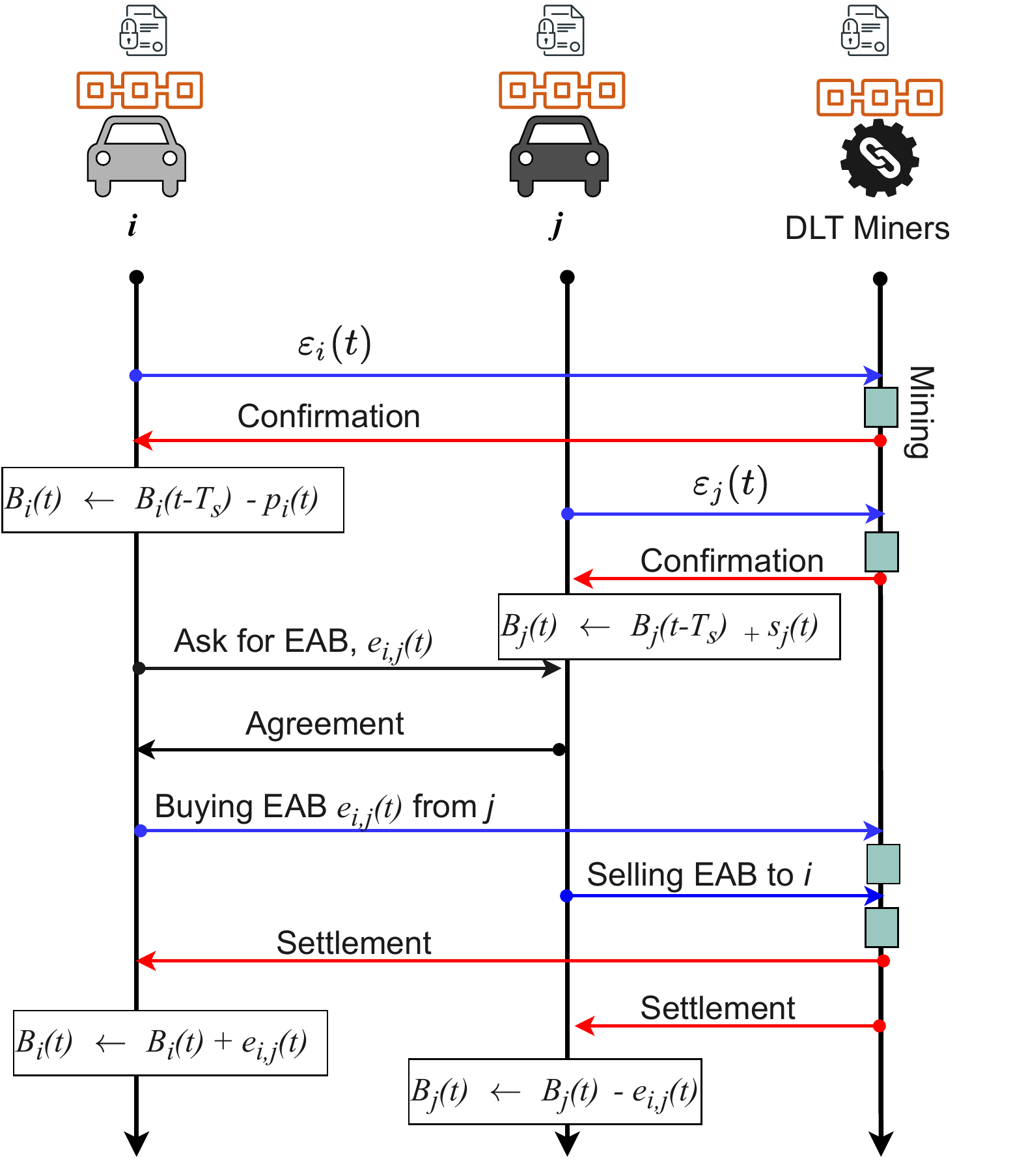}
    \caption{Communication System}
    \label{fig:workflow}
\end{figure}

\subsubsection{Operation}
 The operation of the vehicle's emission allowance trading is performed in the following steps:

\noindent \textbf{Step 1. Publishing Data}. Each vehicle $i\in\mathcal{V}$ computes its own average generated \ch{CO2} emissions, namely, $\epsilon_i(t)$  as shown in Figure \AL{~\ref{fig2systemmodel}} for $t\in\{0,T_s,2T_s,\dotsc,T\}$. These values are published to light ledger version of each vehicle and synchronized with the full ledger stored in DLT full nodes. 
    
\noindent  \textbf{Step 2.  Emission Control}. The generated \ch{CO2} emissions data is recorded in the ledger, and the smart contract with the predefined rules is executed. These rules are characterized by two categories namely maximum \ch{CO2} emissions and actions: warnings, alerts and reminders. 
The published \ch{CO2} data is formatted and arranged into blocks to be verified through a consensus process.
If $\epsilon_i(t) > \mathcal{T}$, the smart contract issues an alert message to $i$ to control its driving behavior and $p_{i}(t)$ is deducted from $B_i(t)$ via smart contract. Hence, the ledger is updated with the value  $B_i(t)\leftarrow B_i(t-T_s) - p_{i}(t)$. In contrast, if $j$ has maintained a safe speed and emitted reasonable amounts of \ch{CO2}, it received an incentive $s_j(t)$ to its balance. Hence, the ledger is updated with $B_j(t)\leftarrow B_j(t-T_s) + s_{j}(t)$.
    
\noindent  \textbf{Step 3. Emission Allowance Trading}. After receiving a confirmation with the required action from the smart contract, if $B_i(t)<0$, then $i$ needs to re-charge its EAB by buying emission allowances from other vehicles. For example, $i$ makes an agreement with $j$ to buy an amount of emission allowances $e_{i,j}(t)$. Then, $i$ sends the buying request for the amount $e_{i,j}(t)$ to execute a smart contract. Next, $j$ updates the smart contract with a selling request and $e_{i,j}(t)$. 
    
\noindent  \textbf{Step 4. Settlement}. Finally, the EAB of each vehicle is updated and settled as $B_i(t) \leftarrow B_i(t)  + e_{i,j}(t)$ and $B_j(t) \leftarrow B_j(t) - e_{i,j}(t)$.

In this paper, we focus on the efficiency of V2V communication between vehicles for exchanging data and trading EAB. We study these in terms of end-to-end latency which includes the transmission latency among vehicles and computation latency of Blockchain validation processes. 

\subsection{Joint Communication and Computation Model}
In this section, we define the total available time for communication between two vehicles and the impact of the Blockchain computation latency. 

\ilm{Let $(x_i(t),y_i(t))$ denote the position of vehicle $i$ at time $t$. If communication is initiated at time $t$, the time in which two vehicles, namely $i$ and $j$, are available for communication is defined by 1) their communication range $r_{ij}$ 2) their positions $(x_i(t),y_i(t))$ and $(x_j(t),y_j(t))$, 3) their relative speed, given by vector $\mathbf{v}_{ij}(t)=\mathbf{v}_i(t)-\mathbf{v}_j(t)$\,km/h. Clearly, to initiate communication at time $t$, the distance between the vehicles must be 
\begin{equation}
    d_{i,j}(t) = \sqrt{(x_i(t) - x_j(t))^2 + (y_i(t) - y_j(t))^2}\leq r_{ij}. 
\end{equation}
Then, the total time for V2V communication between vehicles $i$ and $j$ at time $t$ is given as
\begin{equation}
    L_{total}(t) = \max_{\ell\in\mathbb{R}}\left\{\ell\,\middle|\, d_{i,j}(\ell)\leq r_{ij}\right\}-t.
\end{equation}
 It is immediate to see that $L_{total} \rightarrow \infty $ when $\lVert \mathbf{v}_{ij}(t')\rVert \rightarrow 0$ for all $t'\in\left[t,\ell\right]$. This implies that whenever both vehicles move in the same direction and with near equal speed, they will have a long time $L_{total}$ to communicate and exchange messages. Furthermore, it can be seen that, the upper bound for $L_{total}$\,seconds for the case where the relative speed $\mathbf{v}_{ij}(t')$\,km/h remains constant for all $t'\in\left[t,\ell\right]$ is
 \begin{equation}
    L'_{total}\leq \frac{r_{ij}}{1.8\lVert \mathbf{v}_{ij}(t)\rVert} 
 \end{equation}}
 
\begin{figure}
    \centering
    \includegraphics[width=0.8\linewidth]{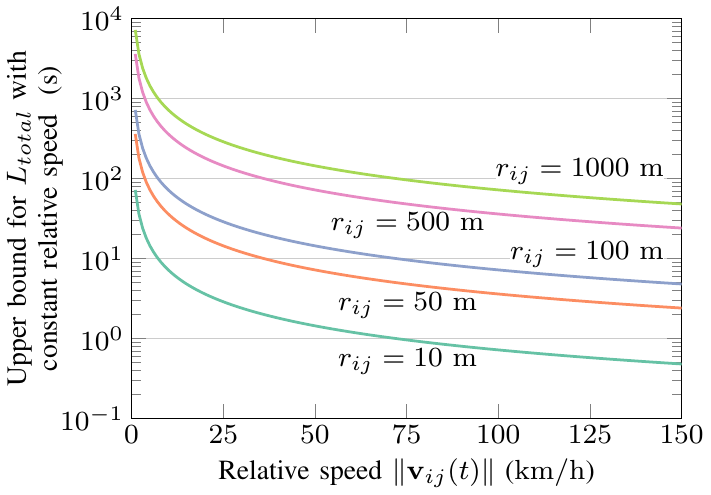}
    \caption{Upper bound of total latency $L_{total}$ for communication between vehicles.}
    \label{fig:le2e}
\end{figure}

 Figure~\ref{fig:le2e} illustrates the upper bound for $L_{total}$ with several values of $\lVert \mathbf{v}_{ij}(t)\rVert$.
 
The time needed to complete a trade between two vehicles $i$ and $j$ in B-ETS can be divided into two parts. First, the communication between vehicles, simply denoted as $L_{trans}$, and, second, the time needed for the verification process in the distributed ledger, denoted as $L_{comp}$. Hence, a trade is completed successfully if and only if
\begin{equation}
    L_{total} \geq L_{trans} + L_{comp}.
\end{equation}
From there, we define the probability of successful data trading as
\begin{equation}
    P_{success} = \Pr\left(L_{comp}  + L_{trans} \leq L_{total} \right)
\end{equation}

The latency for the communication between vehicles $i$ and $j$, denoted simply as $L_{trans}$, is a function of the amount of data that must be exchanged and the effective data rate selected for communication $R$ in packets per second. The data that must be exchanges is defined by the block size of the Blockchain, denoted as $S_B$. On the other hand, the effective data rate $R$ is determined by the implemented protocol, the wireless conditions (e.g., fading, noise, interference, and number of active devices), and the modulation and coding scheme; where the latter determines the instantaneous data rate. The implemented protocol for communication is the IEEE 802.11p standard and the wireless environment are given in Section~\ref{sec:perf_eval}. Nevertheless, we can approximate the latency for communication by assuming that the effective data rate remains constant throughout the trade as 
\begin{equation}
    L_{trans}\approx\frac{S_B}{R}.
\end{equation}

\AL{The formulations to calculate $L_{comp}$ are presented in the following}.

\subsubsection{Blockchain computation latency}
We consider a Blockchain-based VANET network that includes a subset of vehicles $\mathcal{M}\subseteq\mathcal{V}$ that work as miners. These miners start their Proof-of-work (PoW) mechanism computation at the same time and keep executing the PoW process until one of the miners completes the computational task by finding the desired hash value\cite{Bitcoin}. When a miner $i$ executes the computational task for the POW of current block, the time period required to complete this PoW can be formulated as an exponential random variable $W_i$ whose distribution is $f_W(w,i) = \lambda_c e^{-\lambda_c w} $, in which $\lambda_c = \lambda_0 P_c$ presents for the computing speed of a miner, $P_c$ is power consumption for computation of a miner, and $\lambda_0$ is a constant scaling factor. Once a miner completes its PoW, it will broadcast messages to other miners, so that other miners can stop their PoW and synchronize the new block.

For the PoW computation, we are interested in finding the time in which the first miner $i*$, among all the $M=|\mathcal{M}|$ miners, finds out the desired hash value. This is the time for the fastest PoW computation among miners and denoted by the  random variable $W_{i*}$. By assuming $\{W_i\}$ are i.i.d. random variables, we can calculate the complementary cumulative probability distribution of $W_{i*}$ as 
\begin{align}
    \Pr(W_{i*} > w)&= \Pr\left(\min_{i \in \mathcal{M}} (W_i) > w\right)= \prod_{i\in\mathcal{M}} \Pr(W_i > w)\nonumber\\ 
    &= (1 - \Pr(W_i < w))^M, \text{ s.t. } i\in\mathcal{M}.
\end{align}
Hence, $L_{comp}$ is the average computational latency of the fastest miner $i*$, calculated as
\begin{equation}
    L_{comp} = \int_0^\infty (1 - \Pr(W_i \leq w))^M \dd w = \int_0^\infty e^{-\lambda_cMw}\dd w 
\end{equation}
Now we can calculate the communication latency as $L_{trans}+L_{comp}$. 

Note that it can occur that the communication delay exceeds the available communication time $L_{total}$. In such a case, a proposed transactions with potentially valid PoW solution must be abandoned. Hence, finding a valid puzzle solution does not guarantee that the proposed transactions will be finally accepted by the network because of the propagation delay. In such cases, a Blockchain fork can only be adopted as the canonical Blockchain state when it is first disseminated across the network. In scope of this research, to simplify, we do not address the problem of fork, please refer to \cite{wang2019survey} for more detail.    

\section{ \uppercase{Performance Evaluation}}
\label{sec:perf_eval}
  In this section, we analyze the performance of our proposed B-ETS system. 

\begin{table}[t!]
\centering
\caption{SMART CONTRACT EXECUTION COSTS}
 \begin{tabular}{@{}  p{2.5cm}  p{1.2cm} p{1.2cm} p{1.2cm}  @{}}
 \toprule
\textbf{Smart Contracts} & \hspace{0.1cm} \textbf{Gas} & \hspace{0.1cm} \textbf{Ether} & \hspace{0.1cm} \textbf{USD} \\ [0.1ex] 
 \midrule
UserAuthority   & 159430 & 15.9$\cdot 10^{-5}$ & 0.0723  \\ 
RecordData      & 152443 & 15.2$\cdot 10^{-5}$ & 0.0692 \\ 
AlertControl    & 213924 & 21.3$\cdot 10^{-5}$& 0.0971 \\ 
Incentive       & 224934 & 22.4$\cdot 10^{-5}$& 0.1021 \\ 
RecordData      & 276394 & 27.6$\cdot 10^{-5}$& 0.1254 \\ 
EABTransfer     & 246374 & 24.6$\cdot 10^{-5}$& 0.1118 \\ 
\bottomrule
\end{tabular}
    \begin{tablenotes}
    \centering
      \small
      \item * 1 Ether = $10^9$ Gwei; 1 USD = 4,182,471.9949 Gwei
    \end{tablenotes}
\label{tab:gas}
\end{table}

In order to emulate a realistic vehicle network as presented in Figure \ref{fig2systemmodel}, a combination of micro simulators, network libraries and open-source vehicular network simulators is employed. Specifically, SUMO \cite{SUMO1}, OMNET++ which runs in parallel via a proxy TCP connection, and Veins. The IEEE 802.11p standard is used for communication between vehicles and a simple path loss model is selected. In each simulation, 120 vehicles are generated and located randomly. The \ch{CO2} emissions are calculated reading the Traffic Control Interface (TraCI) commands from SUMO. Ethereum is deployed as a ledger in the experiments by using local Ganache platform.


\begin{figure}[t]
\centering
\begin{subfigure}{0.4\textwidth}
\RaggedRight
    \includegraphics[width=0.95\linewidth]{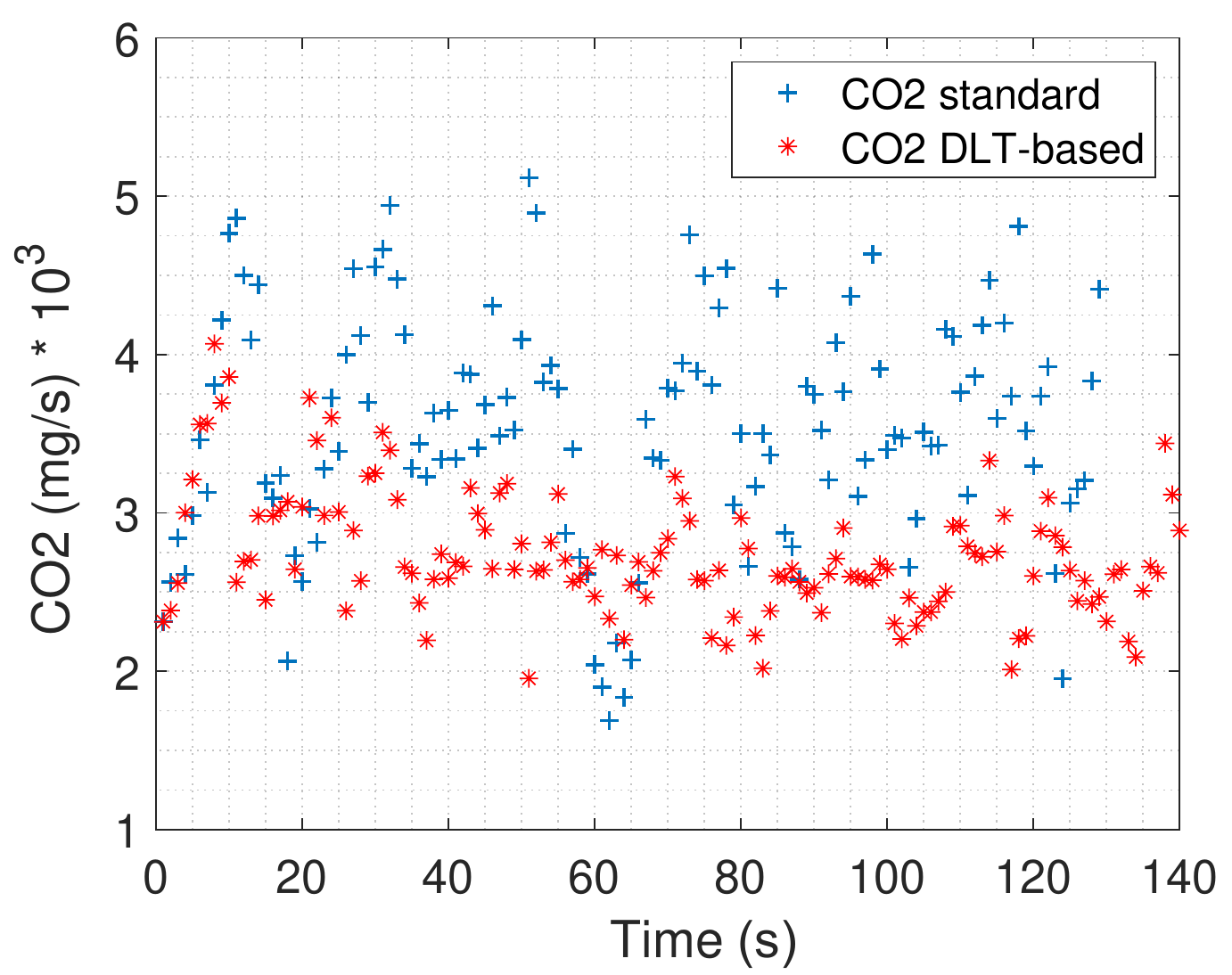} 
    \caption{CO2 Emission}   
    \label{fig:co2}%
\end{subfigure}%
\begin{subfigure}{0.4\textwidth}
\centering
    \includegraphics[width=0.95\linewidth]{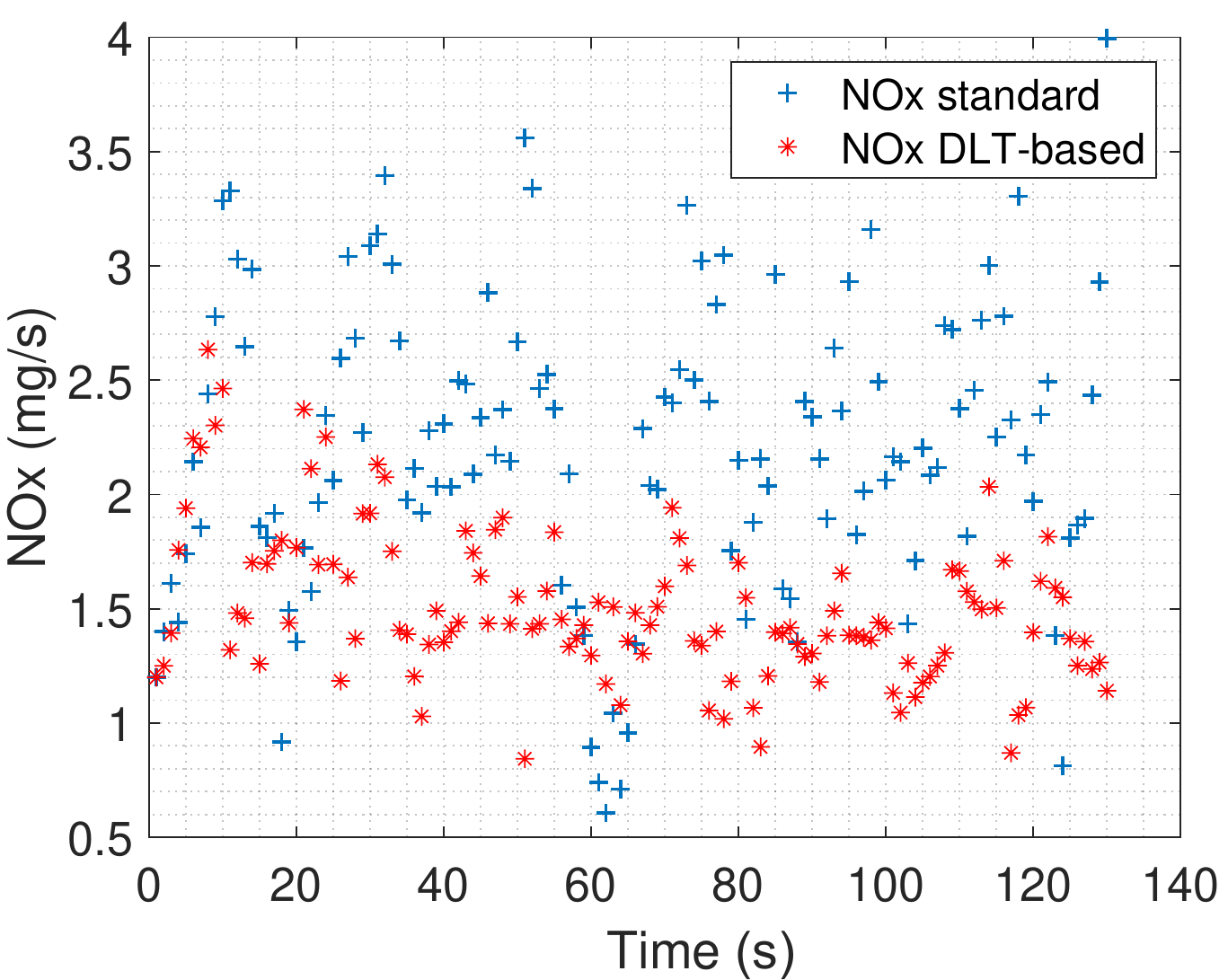}
    \caption{NOx Emission}
    \label{fig:nox}
\end{subfigure}

\vspace{0.3cm}
\begin{subfigure}{0.4\textwidth}
\centering
    \includegraphics[width=0.95\linewidth]{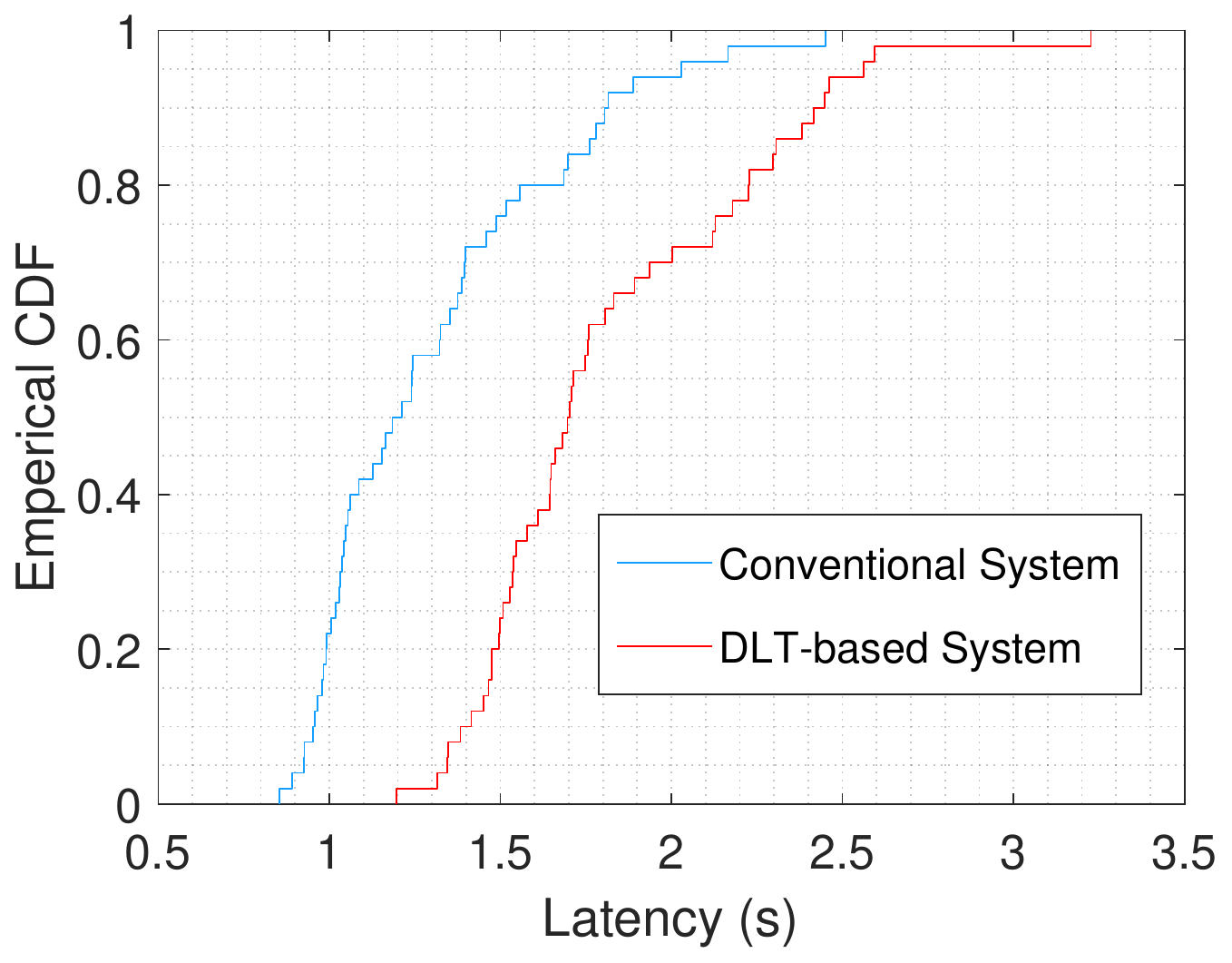}
    \caption{V2V communication latency}
    \label{fig:latency}
\end{subfigure}
\vspace{0.3cm}

\caption{Performance Evaluation. \textit{(a) and (b): The \ch{CO2} and \ch{NOx} emission generated in standard and DLT based systems; (c) Communication latency between standard and Blockchain-based system.}}
\label{fig:results}
\end{figure}

The computational efforts to execute smart contracts in Blockchain are measured in units of gas. The currency for Ethereum is Ether (ETH). In our simulations, the transaction costs and execution costs are converted to ETH and USD, see Table~\ref{tab:gas}. 
The ETH gas station was used to estimate the costs, the price is generated using a static average of 20 Gwei, where one Ether is equivalent to $10^{9}$ Wei. The transaction costs are the costs associated with sending the contract codes to the Ethereum blockchain, dependent on the size of the contract. 

The amount of \ch{CO2} generated from vehicles is dependent upon various factors such as: speed, age of vehicles, etc. We ran two separate experiments to compare the amount of emissions generated between a standard CAP system and a Blockchain-based system when the driving behavior is controlled. Figure~\ref{fig:results} illustrates the generated \ch{CO2} and \ch{NO_x}, along with the V2V communication latency for the standard and the DLT-based trading. In the DLT-based trading, vehicles follow defined rules such as dropping their speed in the smart contract. In Figure~\ref{fig:results} we observe that the amount of \ch{CO2} and \ch{NO_x} generated from DLT-based system is lower than conventional system. These results prove that our system has the ability to reduce the overall \ch{CO2} emitted from vehicles on the network.

In B-ETS, the transactions exchanged between vehicles are encrypted, and verified before attached in the distributed ledger. Therefore, the trusted recording and trading data is guaranteed in comparison with standard system. However, because of extra verification steps in Blockchain, the time to complete a transaction between vehicles is higher. This is a trade-off between trust and latency in Blockchain-based systems.
%



\section{\uppercase{Conclusion}}

\noindent  In this paper, we first proposed a Blockchain-based Emission Trading System, called B-ETS, to support the accounting and monitoring of emissions in vehicular networks. B-ETS provides a trustworthy and transparency for accounting the emissions generated from vehicles. The vehicles can exchange their emission allowances through autonomous smart contracts in a trusted manner. We introduce an economic incentive scheme based on smart contracts to encourage drivers to behave in environmentally friendly ways.

This work provides a mechanism for policy makers, vehicle manufacturers and the EU-ETS to enforce the carbon emissions regulations in a more efficient, secure manner as well as to perform full life-cycle analysis of vehicles. Using the proposed method could result in vehicle manufacturer savings, ensuring that they are not subject to excess emissions fees at the end of the year through the continuous monitoring and reporting of \ch{CO2}.

\AL{The next stage of this work involves further analysis of the current system in two ways. First, we will include the analysis of more pollutants such as Particulate Matter (\ch{PM_x}), Carbon Monoxide (CO), Sulfur Dioxide (\ch{SO_2})} into B-ETS. Then, we will address the limitations of this work by diversifying the vehicles on the network, thereby incorporating other types of vehicles (other than passenger vehicles), such as: buses, vans and trucks.  

\section*{\uppercase{Acknowledgements}}
\noindent This work has been in part supported by the European Union's Horizon 2020 program under Grant 957218 IntellIoT, the Independent Research Fund Denmark (DFF) under Grants Nr. 8022-00284B (SEMIOTIC) and Nr. 9165-00001B (GROW), and the National Science Foundation Graduate Research Fellowship under Grant DGE-1839285.

\bibliographystyle{IEEEtran}
\bibliography{MAIN}

\end{document}